%% file: main.tex
\documentclass[a4paper]{jpconf}
\pdfoutput=1

\usepackage[utf8]{inputenc}
\usepackage[left=2.55cm, right=2.55cm, top=2.55cm, bottom=2.55cm]{geometry}
\usepackage{amsmath,amssymb,amsbsy}
\usepackage{slashed}
\usepackage{xcolor}
\usepackage{graphicx}
\usepackage{url}
\usepackage{cancel}
\usepackage{cite}
\usepackage[colorlinks=true,allcolors=darkpurple,pdfborder={0 0 0},linktocpage=false]{hyperref}
\usepackage{tabularx,booktabs}
\usepackage{multicol}
\usepackage{feynmp}
\usepackage{units}
\usepackage{xspace}
\usepackage[labelfont=bf]{caption}
\usepackage[section]{placeins}
\usepackage{subcaption}
\usepackage{soul} 

\definecolor{darkred}{rgb}{0.6,0,0}
\definecolor{darkpurple}{rgb}{0.5,0,0.5}

\def\hc{\text{h.c.}}

\def\z2{$\mathbb{Z}_2$}
\def\321{$\mathrm{SU(3)_c} \times \mathrm{SU(2)_L} \times \mathrm{U(1)_Y}$}
\def\one{\ensuremath{\mathbf{1}}}
\def\two{\ensuremath{\mathbf{2}}}
\def\three{\ensuremath{\mathbf{3}}}
\def\threeS{\ensuremath{\mathbf{\bar 3}}}
\providecommand{\abs}[1]{\lvert#1\rvert} 

\definecolor{vdrgreen}{rgb}{0.0, 0.7, 0.0}
\definecolor{avblue}{rgb}{0.0, 0.0, 0.8}

\begin{document}


\title{
Dark matter in a charged variant of the Scotogenic model} 
\vspace{0.5cm}

\author{Miguel Puerta} 
\address{ Instituto de F\'{i}sica Corpuscular, CSIC-Universitat de Val\`{e}ncia, 46980 Paterna, Spain}
\ead{miguel.puerta@ific.uv.es}

\begin{abstract}
Scotogenic models are among the most elegant and economic solutions that provide an explanation for two of the main open questions in particle physics: neutrino masses and dark matter (DM). In this work, after a brief discussion of the model, we present a phenomenological study of the DM candidate in a variant of the Scotogenic model. While in the original Scotogenic scenario the DM candidate can be fermionic or bosonic, only the latter is viable in this version. The presence of new charged states might reveal new regions in the parameter space compatible with current observations.
\end{abstract}

\input{tex/intro}

\section*{Acknowledgements}
This paper is based on the talk given at TAUP 2021, available \href{https://www.youtube.com/watch?v=NSfLY7lhigc&t=5s}{here}. The original work \cite{DeRomeri:2021yjo} was done in collaboration with V. De Romeri and A.Vicente. I would like to thank them for their help writing this manuscript. Work supported by the Spanish grants FPA2017-85216-P
(MINECO/AEI/FEDER, UE), SEJI/2018/033, SEJI/2020/016 (Generalitat Valenciana) and FPA2017-90566-REDC (Red Consolider MultiDark).



\section*{References}
\bibliographystyle{iopart-num}
\bibliography{refs}

\end{document}

%% file: tex/intro.tex
\section{Introduction}
\label{sec:intro}
Among other open questions, the Standard Model (SM) of particle physics is not able to explain the origin of neutrino masses and the nature of DM. In spite of these questions not being necessarily linked, it is tempting to explore extensions that can account for both. 

One of these examples is the Scotogenic model \cite{Ma:2006km}. This is an economical setup which only requires the addition of a scalar doublet $\eta$, three singlet fermions and a new $\mathbb{Z}_2$ symmetry (under which all new states are odd whereas all SM ones are even). Hence, the lightest $\mathbb{Z}_2$-odd state would be stable, and, if neutral, constitutes a good DM candidate. On the other hand, the $\mathbb{Z}_2$ protects neutrinos from obtaining a mass at tree-level. Therefore, neutrino masses are radiatively generated (at one-loop). In this version, proposed by Aoki et al. \cite{Aoki:2011yk} (discussed in detail in section 2 of \cite{DeRomeri:2021yjo}), we further extend the scalar sector with the addition of another $SU(2)_L$ doublet $\phi$ with hypercharge ($Y$) 3/2. The three doublets of the scalar sector can be decomposed as:
{
	\renewcommand{\arraystretch}{1.6}
	\begin{table}[t]
		\centering
		\begin{tabular}{ c | c c c c c c c | c c c }
			\toprule
			& $q_L$ & $u_R$ & $d_R$ & $\ell_L$ & $e_R$ & $\psi_L$ & $\psi_R$ & $H$ & $\eta$ & $\Phi$ \\ 
			\hline
			$\rm SU(3)_C$ & $\three$ & $\threeS$ & $\threeS$ & $\one$ & $\one$ & $\one$ & $\one$ & $\one$ & $\one$ &$\one$ \\
			$\rm SU(2)_L$ & $\two$ & $\one$ & $\one$ & $\two$ & $\one$ & $\one$ & $\one$ & $\two$ & $\two$ & $\two$ \\
			$\rm U(1)_Y$ & $\frac{1}{6}$ & $\frac{2}{3}$ & $-\frac{1}{3}$ & $-\frac{1}{2}$ & $-1$ & $-1$ & $-1$ & $\frac{1}{2}$ & $\frac{1}{2}$ & $\frac{3}{2}$ \\[1mm]
			\hline
			$\mathbb{Z}_2$ & $+$ & $+$ & $+$ & $+$ & $+$ & $-$ & $-$ & $+$ & $-$ & $-$ \\
			\textsc{Generations} & 3 & 3 & 3 & 3 & 3 & 2 & 2 & 1 & 1 & 1 \\
			\bottomrule
		\end{tabular}
		\caption{Particle content of the model. $q_L$, $\ell_L$, $u_R$, $d_R$,
			$e_R$ and $H$ are the usual SM fields.
			\label{tab:content}}
	\end{table}
}

\begin{equation}
H = \begin{pmatrix} H^+ \\ H^0 \end{pmatrix} \, , \quad
\eta = \begin{pmatrix} \eta^+ \\ \eta^0 \end{pmatrix} \, , \quad
\Phi = \begin{pmatrix} \Phi^{++} \\ \Phi^{+} \end{pmatrix} \, .
\end{equation}
Here $H$ is the SM Higgs doublet. If we assume that CP is conserved in the dark sector, we can write: $\sqrt{2}\eta^0= \eta_R + i\eta_I$. These mass eigenstates are the only potentially viable DM candidates. On the other hand, the fermionic sector is composed by two generations of vector-like fermions ($\psi_L,\psi_R$) with $Y=-1$. The model content is shown in table \ref{tab:content}. With these ingredients, the most general lagrangian can be written as:

\begin{equation} \label{modlag}
\mathcal{L}_Y = M_{\psi} \, \overline{\psi}_L \, \psi_R + Y^L \, \overline{\ell_L^c} \, \Phi \, \psi_L + Y^R \, \overline{\ell_L} \, \eta \, \psi_R + \hc \, ,
\end{equation}
where $M_{\psi}$ is a $2 \times 2$ vector-like (Dirac) mass matrix,
which we can take diagonal without loss of generality, while $Y^L$
and $Y^R$ are dimensionless $3 \times 2$ complex matrices. On the other hand, the scalar potential can be written as:

\begin{align}
\mathcal{V} = \, & \mu^2_1 \, \abs{H}^2 + \mu^2_2 \, \abs{\eta}^2 + \mu_\Phi^2 \, \abs{\Phi}^2 + \frac{1}{2} \, \lambda_1 \, \abs{H}^4 + \frac{1}{2} \, \lambda_2 \, \abs{\eta}^4 +\frac{1}{2} \, \lambda_\Phi \, \abs{\Phi}^4 \nonumber  \\
+& \lambda_3 \, \abs{H}^2 \, \abs{\eta}^2 + \lambda_4 \, \abs{H^\dagger\eta}^2 + \rho_1 \, \abs{H}^2 \, \abs{\Phi}^2 + \rho_2 \, \abs{\eta}^2 \, \abs{\Phi}^2 + \sigma_1 \, \abs{H^\dagger\Phi}^2 + \sigma_2 \, \abs{\eta^\dagger\Phi}^2 \label{eq:pot} \\
+& \frac{1}{2} \left[\lambda_5 \, (H^\dagger \eta)^2 + \hc \right] + \left[\kappa \, (\Phi^\dagger H)(\eta H)+ \hc \right] \, , \nonumber
\end{align}
where $\mu_1$, $\mu_2$ and $\mu_\Phi$ are parameters with dimension
of mass and $\lambda_j$ ($j=1,2,3,4,5$), $\lambda_\Phi$, $\rho_1$,
$\rho_2$, $\sigma_1$, $\sigma_2$ and $\kappa$ are dimensionless. One of the most relevant terms is the last one (with $\kappa$ as coupling constant). It does not allow us to define a conserved lepton number, and, additionally, it is crucial for the generation of neutrino masses, as we can see in figure \ref{fig:neumod}. The masses of the potential DM candidates are:
\begin{figure}[!b]
	\centering
	\includegraphics[width=0.5\linewidth]{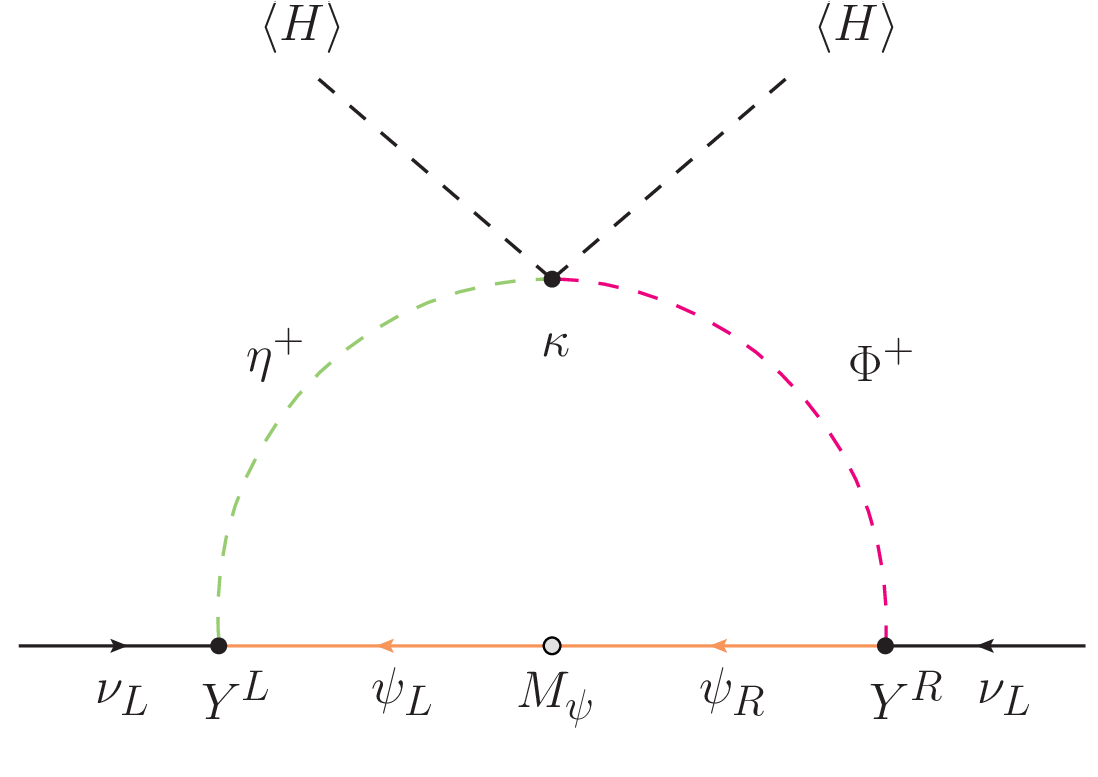}
	\caption{One-loop diagram which generates neutrino masses. In each vertex the relevant components of $\eta$ and $\Phi$ are the singly-charged ones, since the fermions are not neutral.	\label{fig:neumod}	}
\end{figure}
\begin{equation} \label{eq:mEtaRI}
m_{\eta_{R,I}}^2 = \mu_2^2 + \frac{1}{2} \left( \lambda_3 + \lambda_4 \pm \lambda_5 \right) v^2 \, .
\end{equation}
Here $v \simeq 246 \text{ GeV}$. This last expression indicates that the DM candidate is determined by the sign of $\lambda_5$. The neutrino mass matrix expression reads:

\begin{align} \label{eq:mnufinal}
(m_\nu)_{\alpha \beta} = \sum_{b=1}^2 \frac{Y^L_{\alpha b} \, Y^R_{\beta b} + Y^R_{\alpha b} \, Y^L_{\beta b}}{32 \, \pi^2 \, m_{\psi^a}} \frac{\kappa v^2}{m^2_{H^\pm_2}-m^2_{H^\pm_1}}\left(m^2_{H_2^\pm} \log\frac{m^2_{\psi^b}}{m^2_{H_2^\pm}}-m^{2}_{H_1^\pm} \log\frac{m^2_{\psi^b}}{m^2_{H_1^\pm}}\right) \, ,
\end{align}
where $H_1^\pm\text{ and } H_2^\pm$ are the mass eigenstates product of the mixing between $\eta^\pm \text{ and } \Phi^\pm$. If we take the limit $\kappa \to 0$ in (\ref{eq:pot}) and (\ref{eq:mnufinal}),  lepton symmetry is restored and neutrinos become massless, respectively. Additionally, equation (\ref{eq:mnufinal}) manifests that if we choose $Y^L,Y^R \sim 1$, we can generate $m_\nu \sim 1 \text{ eV }$ with $m_{\Psi^a} \sim 1 \text{ TeV, }  m_{H_1} \sim 300 \text{ GeV, } m_{H_1} \sim 400 \text{ GeV }  \text{ and } \kappa \sim 10^{-12}$.

It is important to notice that there are some parameters of the model that cannot be fixed independently: the elements of $Y^L$ and $Y^R$. The general parametrization of Majorana neutrino mass models \cite{Cordero-Carrion:2018xre,Cordero-Carrion:2019qtu} allows us to identify the real degrees of freedom, as well as automatically adjust them to be compatible with neutrino oscillation data~\cite{deSalas:2020pgw} (see appendix B of \cite{DeRomeri:2021yjo}).

\section{Analysis and results}
\label{sec:results}

For a complete description of the analysis we refer to section 3 of \cite{DeRomeri:2021yjo}. The software used for this analysis are \texttt{SARAH (version 4.11.0)}~\cite{Staub:2013tta}, \texttt{SPheno (version 4.0.2)}~\cite{Porod:2003um,Porod:2011nf} (including {\tt FlavorKit}~\cite{Porod:2014xia}) and \texttt{micrOmegas (version 5.0.9)}~\cite{Belanger:2018ccd}. We have performed a scan of $\sim$ 11 000 points in different regions of the parameter space, all of them considering $\eta_R$ as DM candidate (analogous results would have been obtained if we considered $\eta_I$ instead). Each point has been confronted to different constraints, coming from neutrino oscillation data, LHC (including lepton flavor violation (LFV)) and DM searches, among others.

\begin{figure}[!b]
	\centering
	\includegraphics[width=0.7\linewidth]{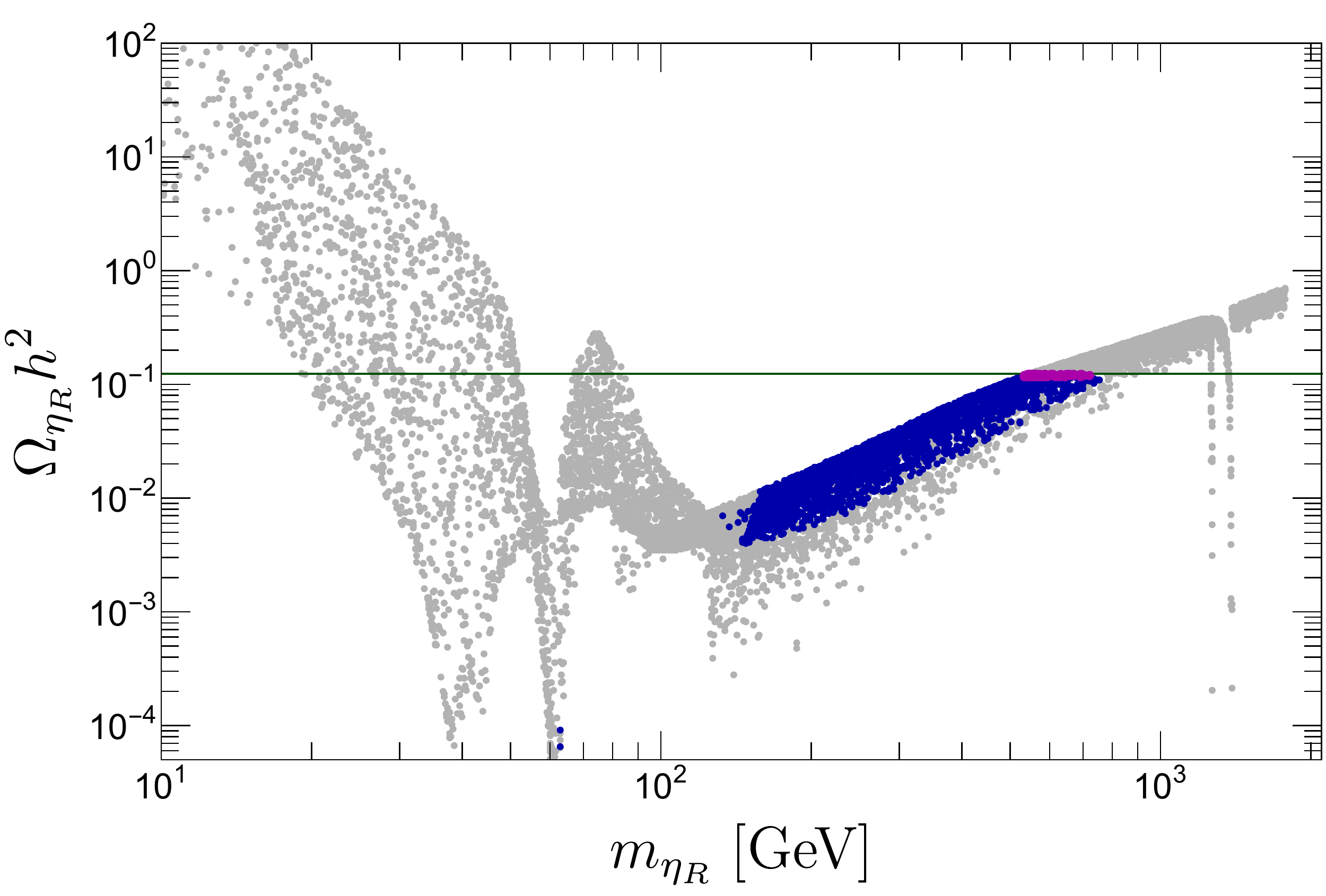}
	\caption{Relic abundance of $\eta_R$ as a function of $m_{\eta_R}$.  Magenta points depict solutions in agreement with the cold dark matter measurement obtained from Planck data~\cite{Aghanim:2018eyx} (green band, 3$\sigma$ interval). Blue points denote allowed solutions but leading to underabundant dark matter. Gray points are excluded by any of the considered constraints.
		\label{fig:omega-metaR}}
\end{figure}

Let us discuss the DM phenomenology. First of all, we have studied in which regions of the parameter space the DM candidate is able to reproduce the observed relic density, as shown in figure \ref{fig:omega-metaR}. As we can see, the preferred region of the points that can account for the total relic density and survive to the different constraints is located at $m_{\eta_R} \sim 500-800 \text{ GeV}$. Furthermore, a very interesting feature occurs at $m_{\eta_R} \sim 1.3 \text{ TeV}$: the relic density suddenly drops to smaller values when the process $H_1^- H_2^+ \to \nu\nu$ becomes efficient. This process requires larger $Y^L,Y^R$ to be determinant, so, there is some conflict between the LFV constraints and this effect. However, it is possible to obtain allowed solutions once some fine-tuning is provided. We point out that this feature depends on other parameters of the model (like the fermion masses). Consequently, this is not an exclusive prediction for this region. One should be able to find it in other regions with other parameter configurations.

\begin{figure}[!hbt]
	\centering
	\includegraphics[width=0.7\linewidth]{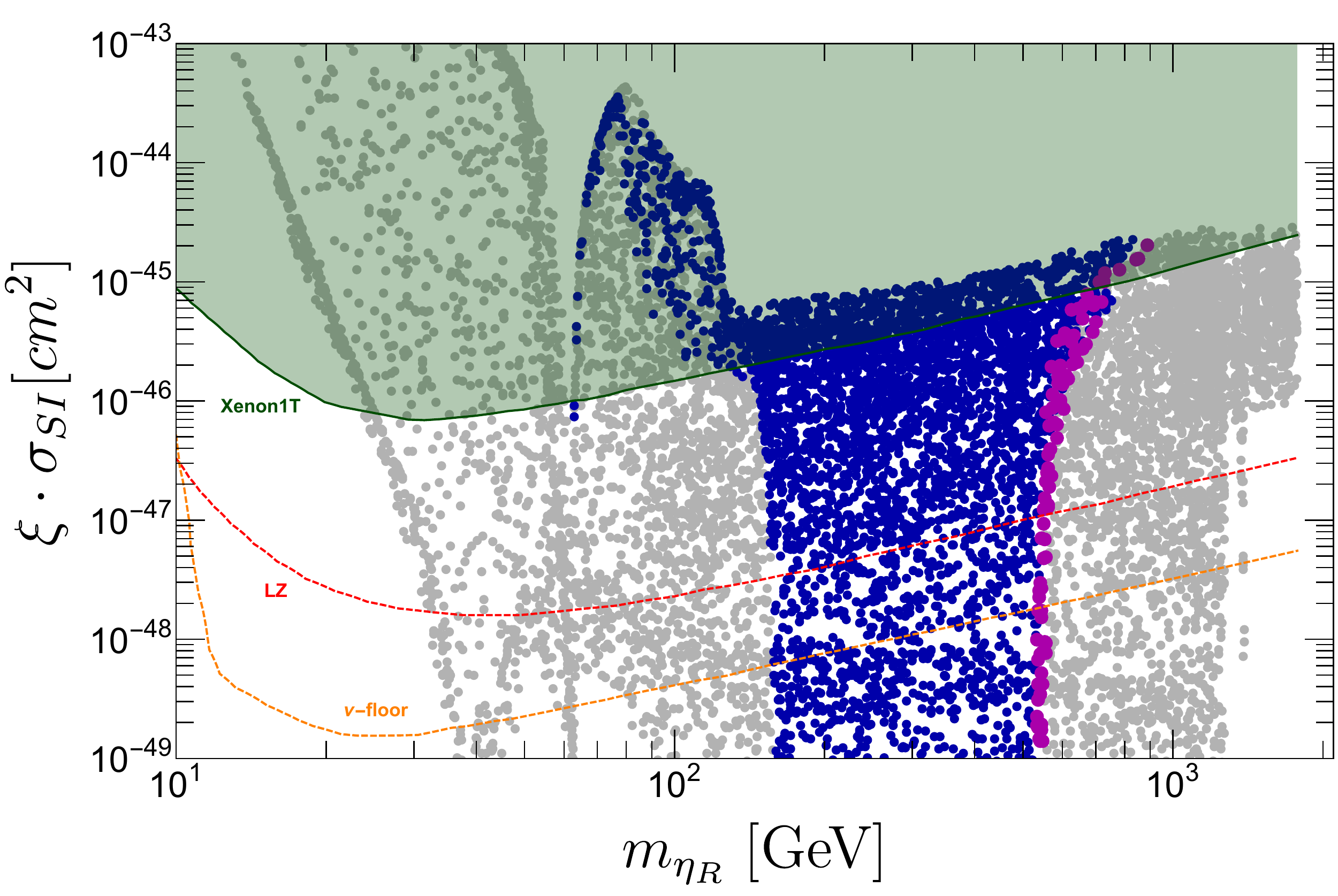}
	\caption{Spin-independent $\eta_R$-nucleon elastic scattering cross section -- weighted by the relative abundance -- versus $m_{\eta_R}$. Same color code as in Fig.~\ref{fig:omega-metaR}. The green dashed area is excluded by the XENON1T experiment~\cite{Aprile:2018dbl}. The dashed orange curve indicates the ``$\nu$-floor" from CE$\nu$NS of solar and atmospheric neutrinos~\cite{Billard:2013qya}. The dashed red curve corresponds to the future sensitivity at LUX-ZEPLIN (LZ)~\cite{Akerib:2018lyp}. 
		\label{fig:DD}}
\end{figure}
\begin{figure}[!hbt]
	\centering
	\includegraphics[width=0.7\linewidth]{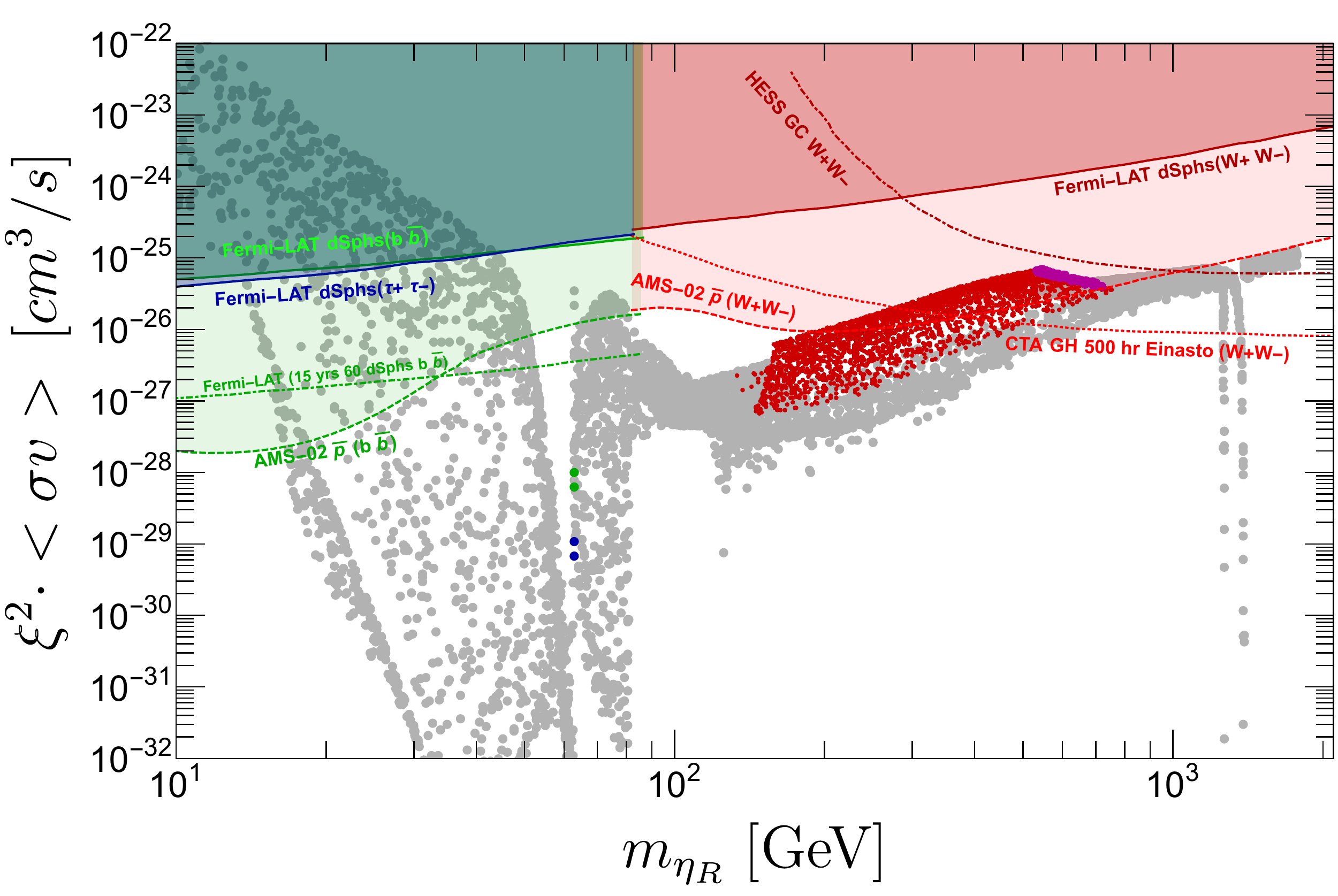}
	\caption{$\eta_R$ annihilation cross section for $b \bar{b}$ (green), $\tau^+ \tau^-$ (blue) and $W^+ W^-$ (magenta when $\eta_R$ makes all the dark matter, red when it would be a subdominant fraction) channels. The green, blue and red plain lines refer to the upper limits currently set by Fermi-LAT $\gamma$-ray data~\cite{Ackermann:2015zua}. The dark red dot-dashed curve is the current obtained by H.E.S.S.~\cite{Abdallah:2016ygi}. The green and red dashed lines denote current constraints derived from the antiproton and B/C data of AMS-02~\cite{Reinert:2017aga}. We also show sensitivity projections for Fermi-LAT ($b \bar{b}$ channel)~\cite{Charles:2016pgz} and for CTA ($W^+ W^-$ channel)~\cite{CTAConsortium:2018tzg}.
		\label{fig:ID}}
\end{figure}
Secondly, we studied its possible future detection. On one hand, the direct detection, by computing the elastic cross section of the $\eta_R$-nucleon interaction. In order to compare it with the current bounds, one should weight it by $\xi = \Omega_{\eta_R}/ \Omega_{CDM}$ (figure \ref{fig:DD}). We can see that the Xenon1T experiment discards some points in the region $m_{\eta_R} \sim 90-200 \text{ GeV}$. However, most of the points easily avoid this constraint. On the other hand, one can study its indirect detection (ID). If $\eta_R$ annihilates into SM particles, its contribution to their astrophysical fluxes might be measurable on Earth. The most suitable candidates to this study are $\gamma$ rays, but other bounds come from antiprotons, for example. As we can see in figure \ref{fig:ID}, solutions which can reproduce the totality of the relic density are in conflict with AMS-02 bounds. We can also see that CTA might be able to explore other regions in the coming future.


%
%
%


\section{Discussion and conclusions}
\label{sec:conclusions}

We have studied in detail the DM phenomenology of a variant of the Scotogenic model, which has additional charged states, including a scalar doublet with $Y=3/2$. Therefore, the model contains a doubly-charged state and several singly-charged ones. This leads to a richer phenomenology than in the Scotogenic model. We have shown that this model correctly reproduces the DM relic density in the same regions of the parameter space as other models, like the Inert Doublet, as well as generating  neutrino masses compatible with the current observations. The allowed points which are able to explain the total DM relic density seem to be in conflict with ID bounds from AMS-02 (figure \ref{fig:ID}). We remark that these bounds have been obtained under significant cosmological uncertainties. Additionally, we found a novel feature (which we can see at $m_{\eta_R} \sim 1.3 \text{ TeV}$ in figures \ref{fig:omega-metaR} and \ref{fig:ID}), which is related to the presence of new charged states. $\Omega_{\eta_R} h^2$ suddenly drops to smaller values when the process $H_1^- H_2^+ \to \nu\nu$ becomes efficient. This requires sizable Yukawa couplings ($Y^L,Y^R$), which lead to a scenario not compatible with some LFV observables. However, some fine-tuning allows to avoid these constraints, enlarging the allowed region of the parameter space. This feature should be found in other regions, providing a novel production mechanism of DM in Scotogenic scenarios.

%% file: main.bbl
\providecommand{\newblock}{}
\begin{thebibliography}{10}
\expandafter\ifx\csname url\endcsname\relax
  \def\url#1{{\tt #1}}\fi
\expandafter\ifx\csname urlprefix\endcsname\relax\def\urlprefix{URL }\fi
\providecommand{\eprint}[2][]{\url{#2}}

\bibitem{Ma:2006km}
Ma E 2006 {\em Phys. Rev. D\/} {\bf 73} 077301 (\textit{Preprint}
  \eprint{hep-ph/0601225})

\bibitem{Aoki:2011yk}
Aoki M, Kanemura S and Yagyu K 2011 {\em Phys. Lett. B\/} {\bf 702} 355--358
  [Erratum: Phys.Lett.B 706, 495--495 (2012)] (\textit{Preprint}
  \eprint{1105.2075})

\bibitem{DeRomeri:2021yjo}
De~Romeri V, Puerta M and Vicente A 2021  (\textit{Preprint}
  \eprint{2106.00481})

\bibitem{Cordero-Carrion:2018xre}
Cordero-Carrión I, Hirsch M and Vicente A 2019 {\em Phys. Rev. D\/} {\bf 99}
  075019 (\textit{Preprint} \eprint{1812.03896})

\bibitem{Cordero-Carrion:2019qtu}
Cordero-Carrión I, Hirsch M and Vicente A 2020 {\em Phys. Rev. D\/} {\bf 101}
  075032 (\textit{Preprint} \eprint{1912.08858})

\bibitem{deSalas:2020pgw}
de~Salas P~F, Forero D~V, Gariazzo S, Mart\'\i{}nez-Mirav\'e P, Mena O, Ternes
  C~A, T\'ortola M and Valle J~W~F 2021 {\em JHEP\/} {\bf 02} 071
  (\textit{Preprint} \eprint{2006.11237})

\bibitem{Staub:2013tta}
Staub F 2014 {\em Comput. Phys. Commun.\/} {\bf 185} 1773--1790
  (\textit{Preprint} \eprint{1309.7223})

\bibitem{Porod:2003um}
Porod W 2003 {\em Comput. Phys. Commun.\/} {\bf 153} 275--315
  (\textit{Preprint} \eprint{hep-ph/0301101})

\bibitem{Porod:2011nf}
Porod W and Staub F 2012 {\em Comput. Phys. Commun.\/} {\bf 183} 2458--2469
  (\textit{Preprint} \eprint{1104.1573})

\bibitem{Porod:2014xia}
Porod W, Staub F and Vicente A 2014 {\em Eur. Phys. J. C\/} {\bf 74} 2992
  (\textit{Preprint} \eprint{1405.1434})

\bibitem{Belanger:2018ccd}
B\'elanger G, Boudjema F, Goudelis A, Pukhov A and Zaldivar B 2018 {\em Comput.
  Phys. Commun.\/} {\bf 231} 173--186 (\textit{Preprint} \eprint{1801.03509})

\bibitem{Aghanim:2018eyx}
Aghanim N {\em et~al.\/} (Planck) 2020 {\em Astron. Astrophys.\/} {\bf 641} A6
  (\textit{Preprint} \eprint{1807.06209})

\bibitem{Aprile:2018dbl}
Aprile E {\em et~al.\/} (XENON) 2018 {\em Phys. Rev. Lett.\/} {\bf 121} 111302
  (\textit{Preprint} \eprint{1805.12562})

\bibitem{Billard:2013qya}
Billard J, Strigari L and Figueroa-Feliciano E 2014 {\em Phys. Rev. D\/} {\bf
  89} 023524 (\textit{Preprint} \eprint{1307.5458})

\bibitem{Akerib:2018lyp}
Akerib D~S {\em et~al.\/} (LUX-ZEPLIN) 2020 {\em Phys. Rev. D\/} {\bf 101}
  052002 (\textit{Preprint} \eprint{1802.06039})

\bibitem{Ackermann:2015zua}
Ackermann M {\em et~al.\/} (Fermi-LAT) 2015 {\em Phys. Rev. Lett.\/} {\bf 115}
  231301 (\textit{Preprint} \eprint{1503.02641})

\bibitem{Abdallah:2016ygi}
Abdallah H {\em et~al.\/} (H.E.S.S.) 2016 {\em Phys. Rev. Lett.\/} {\bf 117}
  111301 (\textit{Preprint} \eprint{1607.08142})

\bibitem{Reinert:2017aga}
Reinert A and Winkler M~W 2018 {\em JCAP\/} {\bf 01} 055 (\textit{Preprint}
  \eprint{1712.00002})

\bibitem{Charles:2016pgz}
Charles E {\em et~al.\/} (Fermi-LAT) 2016 {\em Phys. Rept.\/} {\bf 636} 1--46
  (\textit{Preprint} \eprint{1605.02016})

\bibitem{CTAConsortium:2018tzg}
Acharya B~S {\em et~al.\/} (CTA Consortium) 2018 {\em {Science with the
  Cherenkov Telescope Array}\/} (WSP) ISBN 978-981-327-008-4 (\textit{Preprint}
  \eprint{1709.07997})

\end{thebibliography}
